# Determination of the Solar Rotation Elements and Period from Ruđer Bošković's Sunspot Observations in 1777

*Roša, D.[1], Hržina, D.[1], Husak, M.[2], Brajša, R.[3], Špoljarić, D.[3], Skokić, I.[3], Maričić, D.[1], Šterc, F.[1], Romštajn, I.[1]*

[1] Zagreb Astronomical Observatory, Zagreb, Croatia
[2] Trakošćanska 20, 42000 Varaždin, Croatia, https://orcid.org/0000-0002-2814-5119
[3] Geodetski fakultet Sveučilišta u Zagrebu, Opservatorij Hvar, Fra Andrije Kačića Miošića 26, 10000 Zagreb, Croatia
E-mail: drosa@zvjezdarnica.hr

**Abstract**

This paper focuses on the observations of sunspots made by Ruđer Bošković in 1777. We derived the expressions needed to calculate the elements of the Sun's rotation and period from observations. We used modern ephemeris data in the processing of the observation results. Obtained results are very similar to Bošković's original calculations. In addition to historical significance, they also provide scientifically valuable data on the Sun's differential rotation, which plays a significant role in generating and maintaining solar magnetic activity.

**Introduction**

Ruđer Bošković (1711 - 1787) was one of the most famous Croatian scientists. He was a Jesuit priest and had broad interests and achievements in mathematics, science, civil engineering, philosophy, as well as in poetry and diplomacy (James, 2004). His main research field was astronomy (Špoljarić and Kren, 2016; Špoljarić and Solarić, 2016). Among other things, Bošković developed several methods for the determination of the solar rotation elements and the rotational period of the Sun and applied the methods to his own sunspot observations made in 1777 in Noslon near Sens, 120 km south from Paris (Husak et al., 2021a; Husak et al., 2021b). He was there as a guest of the cardinal Paul d'Albert de Luynes (1703–1788), an amateur astronomer.

Bošković's historical observations are significant because they provide additional data about the solar differential rotation. Many observations show that the solar rotation changes in time (Brajša et al., 1997; Brajša et al., 2006; Wöhl et al., 2010). These changes are in correlation with the solar activity cycle (Jurdana-Šepić et al., 2011; Ruždjak et al., 2017).

Bošković observed sunspots using the telescope with optical micrometre and pendulum for time measurement. In this way it is possible to determine (relatively accurate) the positions of the sunspots on the solar disk. Knowing the coordinates of the centre of the solar disk (e.g. from an astronomical almanac) it is possible to determine the (celestial) equatorial coordinates (declination and right ascension) of the observed sunspot.

As an example, Figure 1 shows a drawing of the solar disk and the position of one sunspot. The line WE passing through the centre of the solar disk is parallel to the plane of the celestial equator. The angular distance of the sunspot from the direction WE corresponds to the difference between the declination of the sunspot ($\delta_f$) and the declination ($\delta_S$) of the centre of the solar disk. The right ascension of the sunspot ($\alpha_f$) is equal to:

$$\alpha_f = \alpha_S + \frac{d}{\cos \alpha_f}, \quad (1)$$

where $d$ is the angular distance of the sunspot from the arc NS which is perpendicular to WE and which passes through the centre of the solar disk and $\alpha_S$ is right ascension of the centre of the solar disk.

From the quantities $\alpha$ and $\delta$, we can easily calculate topocentric ecliptic coordinates of the sunspot (topocentric ecliptic latitude $\beta'_f$ and topocentric ecliptic longitude $\lambda'_f$) according to the well-known transformation formulas between the equatorial and ecliptic coordinate systems:

$$\sin \beta'_f = \cos \varepsilon \sin \delta_f - \sin \varepsilon \cos \delta_f \sin \alpha_f$$
$$\cos \beta'_f \cos \lambda'_f = \cos \delta_f \cos \alpha_f \quad , \text{(2 a, b, c)}$$
$$\cos \beta'_f \sin \lambda'_f = \sin \varepsilon \sin \delta_f + \cos \varepsilon \cos \delta_f \sin \alpha_f$$

where $\varepsilon$ is the obliquity of the ecliptic.

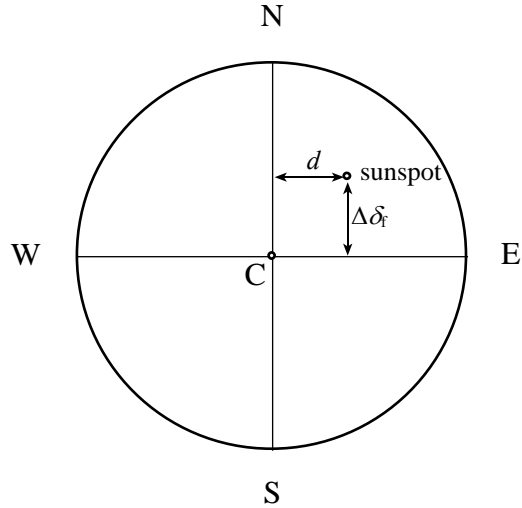

*Fig. 1: drawing of a solar disk with the position of a single spot. Line WE is parallel to the plane of the equator and is easily determined from observations.*

**Expressions needed to find the elements of the Sun's rotation and period from observations**

By the following procedure one can determine the elements of the Sun's rotation and the period from observations. Another approach can be found for e.g. in Waldmeier, 1941. With the known obliquity of the ecliptic $\varepsilon$ at the observation times $t$ and from the expression (2) we can calculate the apparent topocentric ecliptic coordinates of the sunspot ($\lambda'_f, \beta'_f$). The apparent angular and rectangular topocentric coordinates of the Sun (($\lambda'_S, \beta'_S$) and ($x_S, y_S, z_S$)) as well as the current distance to the Sun can be determined using some of the publicly available ephemeris. Due to the small apparent angular distance of the sunspot from the centre of the Sun ($\sigma$) following expression can be used:

$$\sigma = 2 \arcsin \sqrt{\sin^2 \frac{\beta'_f - \beta'_S}{2} + \cos \beta'_f \cos \beta'_S \sin^2 \frac{\lambda'_f - \lambda'_S}{2}}, \quad (3)$$

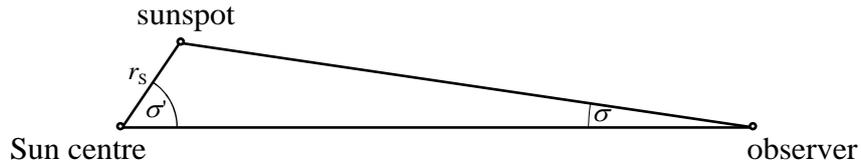

*Fig. 2: with known distance between observer and the Sun ($d_{observer-S}$), apparent angular distance between centre of the Sun and sunspot ($\sigma$) and Solar radius ($r_S$), angle ($\sigma'$) can be determined, as well as heliocentric coordinates of sunspot.*

The heliocentric angular distance between the sunspot and observer is equal to (Fig. 2):

$$\sigma' = 180° - \sigma - \arcsin\left(\frac{d_{observer-S}}{r_S} \cdot \sin \sigma\right), \quad (4)$$

where $d_{observer-S}$ is the topocentric distance to the Sun's centre, and $r_S$ is the radius of the Sun.

The topocentric distance to the sunspot ($d_{f-S}$) is:

$$d_{f-S} = r_S \frac{\sin \sigma'}{\sin \sigma}, \quad (5)$$

The heliocentric rectangular ecliptic coordinates of the sunspot ($x_f$, $y_f$, $z_f$) are:

$$\begin{aligned} x_f &= d_{f-S} \cos \lambda'_f \cos \beta'_f - x_S \\ y_f &= d_{f-S} \sin \lambda'_f \cos \beta'_f - y_S \\ z_f &= d_{f-S} \sin \beta'_f - z_S \end{aligned} \quad (6\ a, b, c)$$

and finally heliocentric ecliptic coordinates of the sunspot (instead of $\lambda_f$, $\beta_f$ in the following text we will use notation $\lambda$, $\beta$):

$$\begin{aligned} \beta &= \arcsin \frac{z_f}{r_S} \\ \lambda &= \arctan \frac{x_f}{y_f} \end{aligned} \quad (7\ a, b)$$

In this procedure we use value of the obliquity of the ecliptic for the instant of time of each observation, i.e. the same that was used during individual transformation of the equatorial coordinates to the ecliptic.

Figure 3 shows the heliocentric celestial sphere. X denotes the position of a sunspot that has coordinates ($\lambda$, $\beta$) in the celestial heliocentric ecliptic system and coordinates ($l$, $b$) in the heliographic system ($b$ is the heliographic latitude, $l$ is arc NX ', where N is the ascending node of the solar equator). Point $P_0$ is the north ecliptic pole, and point $P_S$ is the north pole of the Sun's rotational axis.

The direction of the solar axis in space can be unambiguously determined by two elements: inclination ($i$) - the angle between the ecliptic plane and the solar equatorial plane and ecliptic longitude ($\Omega$) of the ascending node of the solar equator - the angle, in the ecliptic plane, between the equinox direction and the direction where the solar equator intersects (N on Fig. 1) the ecliptic from the South, i.e. in the sense of rotation.

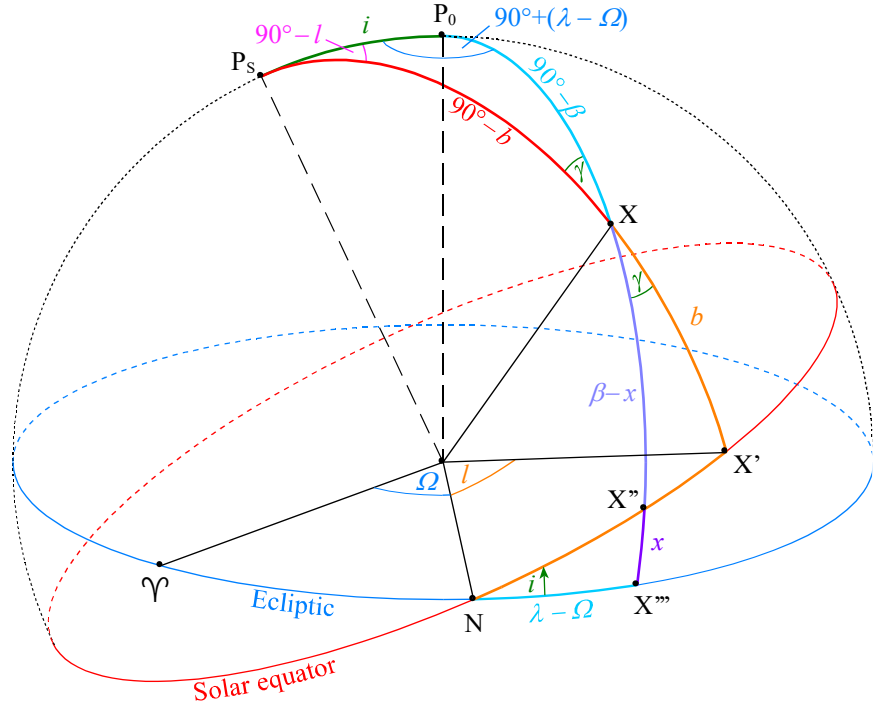

*Fig. 3: coordinates of sunspot* X *on heliocentric celestial sphere in celestial ecliptic system and heliographic system*

By applying the basic formulas of spherical trigonometry to a spherical triangle $P_S X P_0$, we have:

$$\cos i = \cos(90° - b)\cos(90° - \beta) + \sin(90° - b)\sin(90° - \beta)\cos\gamma , \quad (8)$$

from which we find the expression:

$$\cos\gamma = \frac{\cos i - \sin b \sin\beta}{\cos b \cos\beta}. \quad (9)$$

From the triangle XX'X" we have:

$$\cos\gamma = \tan b \cot(\beta - x) = \tan b \frac{1 + \tan\beta \tan x}{\tan\beta - \tan x} \quad (10)$$

and from the triangle NX"X''' we find:

$$\sin(\lambda - \Omega) = \tan x \cot i$$
$$\Rightarrow \tan x = \sin(\lambda - \Omega)\tan i = \sin\lambda\cos\Omega\tan i - \cos\lambda\sin\Omega\tan i . \quad (11)$$

By eliminating the angle $\gamma$ from equations (9) and (10) we can get the expression:

$$\frac{\sin b}{\cos i}\left(\frac{1}{\sin\beta\cos\beta}\right) + \frac{\tan x}{\sin\beta} = \frac{1}{\cos\beta} \quad (12)$$

and by substituting $\tan x$ according to expression (11), we finally find:

$$X + Y\cos\beta\sin\lambda - Z\cos\beta\cos\lambda = \sin\beta , \quad (13)$$

where the $X = \sin b/\cos i$, $Y = \cos\Omega \tan i$ and $Z = \sin\Omega \tan i$ are unknown quantities.

Expression (13) for processing measurements originates from Delambre, (1814). Equation (13) contains three unknowns. So it is necessary to have at least three equations to determine $X$, $Y$ and $Z$. In other words, we need to have at least three observations of a sunspot to get one value of $b$, $\Omega$ and $i$. Of course, we neglect the already small meridional motions of the sunspot, that is, we assume that $b$ is a constant for a particular sunspot. After we determine $X$, $Y$, $Z$ we easily calculate $b$, $\Omega$, $i$ with the following expressions:

$$b = \arcsin(X \cos i)$$
$$\Omega = \arctan\left(\frac{Z}{Y}\right) \quad \quad . \text{(14 a, b, c)}$$
$$i = \arctan\left(\sqrt{Y^2 + Z^2}\right)$$

If we have more than three observations, we can calculate the error of quantity $\Omega$ and $i$ (Stark and Wöhl, 1981).

To calculate the sidereal velocity of the Sun's rotation (and sidereal period of rotation), it is sufficient to determine the change in the quantity $l$ in time. It is possible to derive different formulas for calculating the quantity $l$, but the most suitable is the one in which $l$ is the argument of the tangent function. From a spherical triangle $P_S P_0 X$ (Fig. 3) we find:

$$\cos b \cos l = \cos \beta \cos(\lambda - \Omega)$$
$$\cos b \sin l = \sin \beta \sin i + \cos \beta \cos i \sin(\lambda - \Omega) \quad . \text{(15 a, b)}$$

If we substitute the $\cos b$ from the first formula into the second, we have:

$$\tan l = \frac{\sin i \tan \beta}{\cos(\lambda - \Omega)} + \cos i \tan(\lambda - \Omega) \quad . \text{(16)}$$

From the sidereal period, we can calculate the synodic period by taking into account the details of the Earth's motion (Roša et al., 1995; Skokić et al., 2014).

**Processing of Bošković's measurements**

The individual observation times $t_x$ (for first and second contacts of the solar disk, and for the sunspot) recorded in Bošković's work (Boscovich, 1785) have been measured as the time that elapsed from the moment when Sun was in upper culmination in Sens, France. Therefore, the right ascension is directly given by measured time. Distance in declination from northern edge of the Sun was measured by use of micrometer. The instant of measurement time can be expressed as $t_{xUT1} = t_{s\ kulmUT1} + t_x$. There where 6 days of observations (between 12 and 19 September 1777) of the same sunspot with set of 5 measurements each day of observation. Let the observation time $t$ be the arithmetic mean of the first $t_{1UT1}$ and the last moment of observation $t_{nUT1}$. Also, $\Delta\alpha_f$ and $\Delta\delta_f$ are the mean distances from the centre of the solar disc in right ascension and declination respectably for a given set of measurements. Then the coordinates of the sunspot at the moment of observation ($t$) are the difference between the apparent topocentric equatorial coordinates of the Sun ($\alpha_s$; $\delta_s$) and the observed angular distances of the sunspot from the centre of the solar disc in right ascension and declination respectively ($\Delta\alpha_f$, $\Delta\delta_f$). Heliocentric ecliptic coordinates of the sunspot ($\lambda$, $\beta$) are then obtained by using expressions (2 a, b, c), (3), (4), (5), (6 a, b, c) and (7 a, b). By a combination of coordinates obtained from three observations and solving system of three equations (13) we calculate $b$, $\Omega$, $i$ using expressions (14 a, b, c) for each triplet of observations. Mean values of $b = 26.43°$, $\Omega = 65.61°$, $i = 7.88°$ are used for calculation of $l$ (expression 16) for each of 6 sets of measurements. Result for sidereal period at mean latitude $b = 26.43°$ using least square method is $P = (26.696 ± 0.044)$ days ($\omega = (13.485 ± 0.022)$ °/day) which corresponds to mean value of the synodic period of $P = (28.780 ± 0.047)$ days ($\omega = 12.508 ± 0.020$ °/day). In calculation of the sidereal period from synodic we used mean angular heliocentric velocity of the Earth (0,9765°/day) during period between 12 and 19 September 1777. Bošković's results for sidereal period is 26.77 days and it is close to our result.

## Conclusions

This paper presents an analysis of a part of Bošković's observations of sunspots with a presented mathematical approach and with the use of modern ephemeris parameters. The obtained values for the solar rotation parameters and the rotation period are close to Bošković's results, which indicates the accuracy of Bošković's methods. Furthermore, we plan to extend the processing to all of Bošković's observations and compare them with the results of other historical and modern measurements.

## Acknowledgements

R. Brajša and I. Skokić acknowledge the support by the Croatian Science Foundation under the project 7549 "Millimeter and submillimeter observations of the solar chromosphere with ALMA".